# Enhancing Unconventional Spin-Orbit Torque Efficiency: Numerical Study on the Influence of Crystallographic Texture and Polycrystalline Effects on Low-Symmetry Materials

Yifei Yang[1], and Jian-Ping Wang[1]

[1] Department of Electrical and Computer Engineering, University of Minnesota, Minneapolis, MN 55455, USA

**Spin-orbit torque (SOT) has been extensively studied as a key mechanism in spintronics applications. However, conventional SOT materials limit the spin polarization direction to the in-plane orientation, which is suboptimal for efficient magnetization switching. Recently, spin currents with spin polarization along multiple directions have been observed in low-symmetry materials, offering a promising energy-efficient strategy for the field-free switching of magnetic materials with perpendicular magnetic anisotropy. However, the efficiency of this mechanism is highly dependent on the crystallographic texture of the SOT materials, a critical factor that, to date, has not been quantitatively investigated. In this study, we present a comprehensive numerical investigation into the impact of both in-plane and out-of-plane crystallographic textures of SOT materials on the unconventional SOT generated by Dresselhaus-like and out-of-plane spin polarizations. By employing a theoretical orientation distribution function, we calculate the effective unconventional SOT values for SOT materials with tunable crystallographic texture. This analysis provides a framework for the synthesis and optimization of future low-symmetry SOT materials, which can enhance operational efficiency for spintronics applications in magnetoresistive random-access memory and spin logic devices.**

## I. Introduction

Magnetic tunnel junction (MTJ) consisting of two ferromagnetic layers (FMs) separated by an insulating barrier serves as the fundamental building block for magnetoresistive random-access memory (MRAM) and spin logic devices [1], [2]. Spin-orbit torque (SOT) has emerged as an efficient mechanism for manipulating the magnetization of the free FM layer, which encodes the stored information in the MTJ. Notably, magnets with perpendicular magnetic anisotropy (PMA) are favored due to their superior scalability for memory and logic applications compared to in-plane magnets [3], [4]. However, switching of the magnetization of a magnet with PMA requires the assistance of an external magnetic field, which has been an obstacle for applications of SOT-based devices. To overcome this limitation, various approaches have been proposed and demonstrated, including asymmetric geometry [5], tilted anisotropy [6], exchange coupling between FM and antiferromagnet (AFM) [7], stray field from FM [8], FM/non-magnet/FM trilayer structure [9] and combination of SOT and spin-transfer torque [10].

Recently, unconventional SOT generated by spin currents with out-of-plane (OOP) polarization direction, have been discovered in low-symmetry materials [11]-[18], the magnitude of which is dependent on the unconventional spin Hall conductivity (USHC). Spin Hall conductivity tensor $\sigma_{jk}^i$ has three indices, each of which can take either $x$, $y$ or $z$. This SHC tensor describes the amount of spin current with polarization along $i$, flowing along $j$, induced by a charge current along $k$ [19]. The efficiency of switching a PMA magnet is directly proportional to SHC. Heavy metals and topological materials are commonly used SOT materials. However, due to their high symmetry, only conventional SHC elements are allowed, where the three indices are all different, such as $\sigma_{zx}^y$. Recently, SOT materials with low symmetry are found to exhibit USHC, where the spin current's flow direction and polarization direction are the same, such as $\sigma_{zx}^z$. These USHC can generate OOP spin, which has spin polarization direction normal to the surface of the SOT material. OOP spins have been shown to be significantly more efficient for switching PMA magnets than conventional SHC [20], [21]. As a result, significant efforts have been devoted to discovering materials with high USHC values.

SOT thin films are commonly grown using sputtering on polycrystalline or amorphous buffer layers, which often results in a lack of preferred texture [22]-[25]. On the other hand, thin films can be engineered to form a texture, with a minimal mosaic spread of orientations around the preferred axis [12], [14]. While field-free switching can be realized by USHC in low-symmetry SOT materials, its effectiveness is constrained by the crystallographic texture of the materials. Although the impact of domain orientations on the USHC has been discussed qualitatively [12], [13], a systematic and quantitative investigation has not yet been conducted. Here we systematically examine the influence of the in-plane and out-of-plane crystallographic texture on the USHC values of low-symmetry SOT materials. We find that OOP spins are entirely cancelled out if the material is polycrystalline without a preferred orientation. We also calculate the effective SHC values for a textured SOT material, using Mn$_3$GaN [12] as an example, based on the Gaussian distribution function of the texture. Additionally, we summarize approaches to improve crystallographic texture. This analysis provides guidance for the future development of SOT devices utilizing low-symmetry materials.

## II. RESULTS

SHC tensor is given with respect to the crystal axes of materials ($x$, $y$ and $z$). However, materials grown by sputtering are typically polycrystalline, containing grains with various orientations. This introduces a gap between the intrinsic SHC and the experimentally observed SHC in a device. In experiments, axes are defined with respect to the device, which we denote as $X$, $Y$ and $Z$. In this study, we first focus on the in-plane orientation of the SOT material grains, assuming the OOP axis of the crystal aligns with that of the device. Then we studied the influence of OOP texture on USHC.

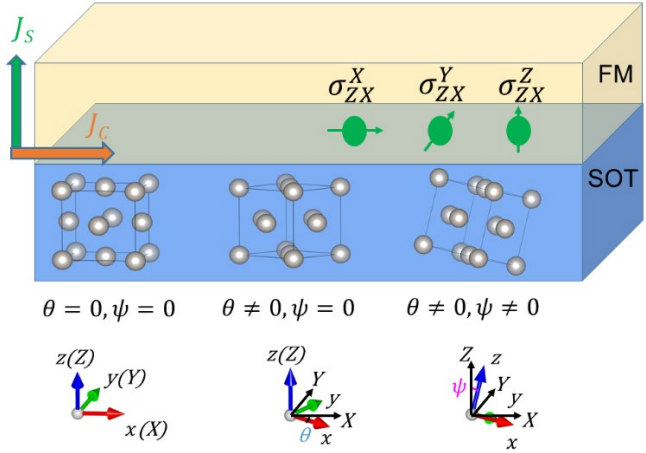

Fig. 1. Schematic of a SOT device composed of a SOT layer and an FM layer. The grains in the SOT layer exhibit various in-plane and out-of-plane orientations represented by $\theta$ and $\psi$, respectively. The crystal uses the coordinate system ($x$, $y$ and $z$) whereas the device uses ($X$, $Y$, $Z$). $\theta$ is the in-plane rotation angle around the $z$ ($Z$) axis, and $\psi$ is the out-of-plane rotation angle around the in-plane rotated $x$ axis. Spin currents will be generated with polarization directions along $X$, $Y$ or $Z$, evaluated by device SHC components $\sigma_{ZX}^X$, $\sigma_{ZX}^Y$, and $\sigma_{ZX}^Z$, which are simplified as $\sigma^X$, $\sigma^Y$ and $\sigma^Z$, respectively.

As shown in Fig. 1, firstly, the crystal axes $x$ and $y$ rotate in-plane around the common $z$ ($Z$) axis by an angle of $\theta$. The $X$ direction is defined by the charge current direction in device. Then, the crystal rotates around the new $x$ axis by an angle of $\psi$, which is an OOP rotation. Since a device can contain grains with various orientations, the effective SHC values of a device are influenced by the orientation distribution of these grains. We study the three SHC tensors depicted in Fig. 1 for the device, with spin current flowing along $Z$ and charge current flowing along $X$, which are the only elements that contribute to the FM switching. Among these three, $\sigma_{ZX}^Y$ is the conventional SHC, while $\sigma_{ZX}^Z$ is unconventional that generates OOP spins. The two sub-indices $Z$ and $X$ will be omitted for simplicity hereafter.

TABLE I
SHC elements for Pt and Mn$_3$GaN. Unit is $\frac{\hbar}{2e}\frac{1}{\Omega\text{cm}}$

| Pt | | Mn$_3$GaN | | |
|---|---|---|---|---|
| $\sigma_{zx}^x = 0$ | $\sigma_{zy}^x = -2750$ | $\sigma_{zx}^x = 29$ | $\sigma_{zy}^x = -114$ | $\sigma_{yx}^y = -45$ |
| $\sigma_{zx}^y = 2750$ | $\sigma_{zy}^y = 0$ | $\sigma_{zx}^y = 114$ | $\sigma_{zy}^y = -29$ | $\sigma_{yz}^z = 29$ |
| $\sigma_{zx}^z = 0$ | $\sigma_{zy}^z = 0$ | $\sigma_{zx}^z = 45$ | $\sigma_{zy}^z = -45$ | $\sigma_{zz}^y = 40$ |

We will first study the influence of in-plane texture on SHC. For grains with certain unique $\theta$, which is the in-plane orientation with respect to $X$, using the projection between crystal axes and device axes, one can get

$$\sigma^X(\theta) = \sigma_{zx}^x \cos^2\theta + \sigma_{zy}^y \sin^2\theta + \left(\sigma_{zy}^x + \sigma_{zx}^y\right)\cos\theta\sin\theta \quad (1)$$

$$\sigma^Y(\theta) = \sigma_{zx}^y \cos^2\theta - \sigma_{zy}^x \sin^2\theta + \left(\sigma_{zy}^y - \sigma_{zx}^x\right)\cos\theta\sin\theta \quad (2)$$

$$\sigma^Z(\theta) = \sigma_{zx}^z \cos\theta + \sigma_{zy}^z \sin\theta \quad (3)$$

We will examine two spin-orbit torque (SOT) materials as examples: Pt, which exhibits only conventional SHC elements [22], and Mn$_3$GaN, which displays both conventional and unconventional SHC elements [12]. Mn$_3$GaN [12] is selected as the representative unconventional SOT material due to its unique spin transport properties, with 24 out of 27 SHC tensor elements being non-zero, providing the best universality for this study. The relevant SHC elements of the two materials are summarized in Table I. Pt serves as a prototype material for crystals with high symmetries which exhibit only conventional SHC elements and Mn$_3$GaN represents crystals with USHC elements.

For Pt, only SHC with Y spin polarization direction ($\sigma^Y$) is non-zero and remains constant across different $\theta$ angles, while both $\sigma^X$ and $\sigma^Z$ are zero. In contrast, all the three device SHC elements can be non-zero for Mn$_3$GaN, due to its intrinsic unconventional SHC components. It is found that $\sigma^X$ and $\sigma^Y$ have period of 180° whereas $\sigma^Z$ displays a period of 360°. Notably, the average value of $\sigma^Y$ is non-zero while both $\sigma^X$ and $\sigma^Z$ oscillate around zero, resulting in zero average values. This implies that for materials with USHC, if the grain orientations are completely random, the effective USHC for device, $\sigma_{eff}^X$ and $\sigma_{eff}^Z$, will cancel out. However, the conventional SHC for device, $\sigma_{eff}^Y$, will retain a non-zero value even in the absence of a preferred orientation.

To understand the influence of in-plane polycrystalline effects on SHC values, we will further examine the effective

SHC values in a device that incorporates a polycrystalline SOT material with a preferred in-plane orientation. The texture of a polycrystalline material can be quantitatively described by orientation distribution function (ODF) [26]. Here we focus on the in-plane orientation of SOT materials, and the ODF will have only one variable, which is the in-plane angle $\theta$. For materials without a preferred orientation, referred to as randomly textured materials, the ODF takes the form of a uniform distribution. Conversely, materials with a preferred orientation, known as textured materials, can be quantitatively described by an ODF with a Gaussian shape [26], [27], with the form of

$$f(\theta, \theta_0, w_{ip}) = A\, exp\left(-\frac{(\theta-\theta_0)^2}{2\, w_{ip}^2}\right) \quad (4),$$

where A is a normalization constant, $\theta_0$ is the mean angle of the Gaussian distribution, which represents the orientation that most grains adopt and is referred to as the preferred orientation. $w_{ip}$ is the standard deviation of the distribution, measuring how much the grain orientations deviate from the mean value. Gaussian distribution functions with mean value of 0° and three different width values are shown in Fig. 2. A smaller width indicates a stronger in-plane texture in the material, meaning that the grains are more consistently aligned. In contrast, when the width approaches 360°, the distribution resembles a uniform distribution (see the inset of Fig. 2), which is characteristic of a randomly textured material.

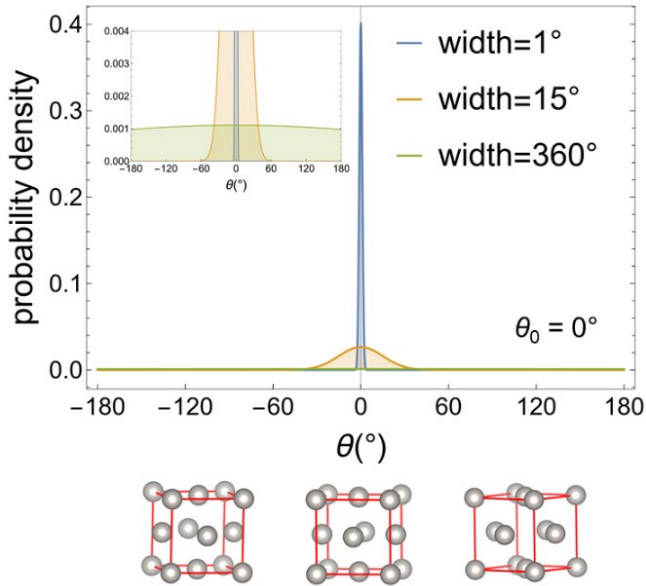

Fig. 2. Illustration of a Gaussian-shape in-plane ODF with $\theta_0$ of 0° and width of 1°, 15° and 360°. The inset shows a zoomed in figure for a better view of the ODF with width of 360°.

As we already have the function of SHC for different in-plane orientations $\theta$ (equations (1)-(3)) and the orientation distribution function shown in Fig. 2, the effective device SHC considering all grains can be computed by

$$\sigma_{eff}(w_{ip}, \theta_0) = \frac{\int \sigma(\theta) f(\theta, \theta_0, w_{ip}) d\theta}{\int f(\theta, \theta_0, w_{ip}) d\theta} \quad (5),$$

where $\sigma(\theta)$ is the orientation-dependent SHC formulated by equations (1)-(3), and $f(\theta, \theta_0, w_{ip})$ is the ODF defined by equation (4). Since the in-plane orientation $\theta$ is the relative angle between the crystal axis $x$ and current injection direction $X$, the mean value of $\theta$ ($\theta_0$) can be set arbitrarily during the device patterning of a textured SOT material. Here we choose one $\theta_0$ value for Pt as its SHC is independent of angle, and three $\theta_0$ values (-45°, 0° and 45°) for Mn$_3$GaN. The effective device SHC is plotted in Fig. 3 as a function of width of the in-plane ODF.

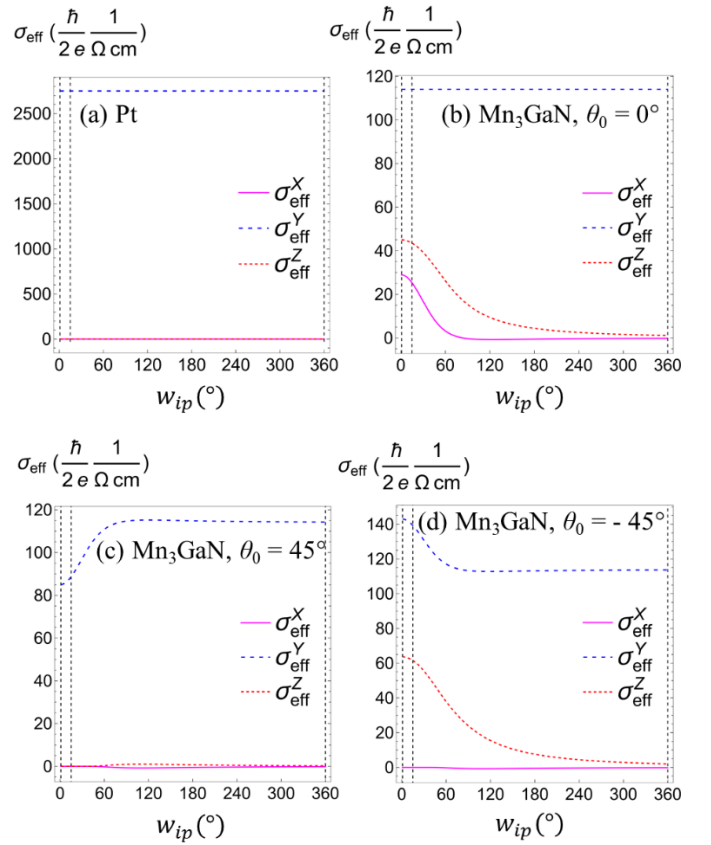

Fig. 3. Computed effective SHC values as a function of width of the in-plane ODF for (a) Pt, and Mn$_3$GaN with $\theta_0$ of (b) 0°, (c) 45° and (d) -45°. The three vertical dashed lines in (b)-(d) indicate width values of 1°, 15° and 360°, from left to right. The calculations are based on SHC values of Pt [22] or Mn$_3$GaN [12].

As shown in Fig. 3(a), the effective SHC for Pt is independent on the width of the Gaussian ODF, which is expected since the device SHC itself is angle-independent. For Mn$_3$GaN, however, the effective SHC depends on the width of the orientation distribution. When the current injection direction $X$ is aligned with the $x$-axis of the SOT material, $\sigma^X$ is maximized, leading to a maximized $\sigma^X_{eff}$ when $\theta_0$ is 0° (Fig. 3(b)). With this $\theta_0$, $\sigma^Y_{eff}$ is independent on width values as $\sigma(\theta)$ is an odd function with respect to $\theta_0 = 0°$. Similarly, $\sigma^Z_{eff}$ can be maximized if $\theta_0$ is -45° (Fig. 3(d)). One common characteristic of Fig. 3(b)-(d) is that the effective device USHC ($\sigma^X_{eff}$ and $\sigma^Z_{eff}$) converge to zero when the width increases. In contrast, the conventional SHC ($\sigma^Y_{eff}$) will converge to a non-zero value, which is consistent with previous discussions about the average SHC.

Next, we analyze the influence of OOP texture on the USHC value, using $\sigma^Z$ as an example. We first calculate the $\sigma^Z$ values for different OOP ($\psi$) and in-plane ($\theta$) orientations. Since it involves two orientations, here we employ the rotation matrix for the calculation. The device SHC and crystal SHC are related by [15]

$$\sigma_{device\,IJ}^{K} = \sum_{l,m,n} D_{Il} D_{Jm} D_{Kn} \sigma_{crystal\,lm}^{n} \quad (6),$$

where $I, J, K$ are the device coordinates that can take either $X$, $Y$ or $Z$. $l, m, n$ are the crystal coordinates that take $x, y$ or $z$. $D$ is the rotation matrix, which has the form of

$$D = \begin{bmatrix} \cos\theta & \sin\theta\cos\psi & \sin\theta\sin\psi \\ \sin\theta & \cos\theta\cos\psi & \cos\theta\sin\psi \\ 0 & \sin\psi & \cos\psi \end{bmatrix},$$

given the definitions of $\theta$ and $\psi$ shown in Fig. 1. Equations (1)-(3) can be reproduced by setting $\psi = 0°$. For textured material, the OOP orientation distribution is symmetric about 0°. Thus, only even components of $\psi$ contribute to the effective $\sigma_{device}$, and odd components cancel out for grains with opposite $\psi$ angles. Therefore, in our calculation, we only consider the components that are even to $\psi$. The SHC for OOP spin, $\sigma^Z$, is expressed as

$$\sigma^Z(\theta,\psi) = \sigma^y_{yx}\cos\theta\sin^2\psi + (\sigma^z_{yz} + \sigma^y_{zz})\sin\theta \sin^2\psi \cos\psi + \sigma^z_{zx}\cos\theta \cos^2\psi + \sigma^z_{zy}\sin\theta \cos^3\psi \quad (7),$$

which is a function of both $\theta$ and $\psi$. The dependence of $\sigma^Z$ on $\psi$ is shown in Fig. 4(a), for $\theta = -45°, 0°, 45°$. Same as our previous discussion, $\theta$ of $-45°$ yields the maximum $\sigma^Z$ value. $\sigma^Z$ can also change sign if the OOP orientation is large.

To include the OOP texture, the ODF function is expanded to:

$$g(\theta,\psi,\theta_0,w_{ip},w_{oop}) = A\, exp\left(-\frac{1}{2}\left[\left(\frac{\theta-\theta_0}{w_{ip}}\right)^2 + \left(\frac{\psi}{w_{oop}}\right)^2\right]\right) \quad (8)$$

Here, $w_{oop}$ is the OOP spread width for the OOP texture distribution, analogous to $w_{ip}$ for in-plane texture. The mean value of $\psi$ is chosen as 0° for the symmetric distribution

consideration. The effective $\sigma^Z$ is then calculated as (with the in-plane spread width of 15°):

$$\sigma^Z_{eff}(w_{oop},\theta_0) = \frac{\iint \sigma^Z(\theta,\psi)g(\theta,\psi,\theta_0,w_{oop})d\theta d\psi}{\iint g(\theta,\psi,\theta_0,w_{oop})d\theta d\psi} \quad (9)$$

The $\sigma^Z_{eff}$ values for $\theta_0 = -45°, 0°, 45°$ are shown in Fig. 4(b). It is evidence that the effective SHC for Z-polarized spins degrades as the OOP spread width increases, reflecting worse OOP texture. Again, $\theta_0 = -45°$ leads to the highest $\sigma^Z_{eff}$. Compared with in-plane texture, $\sigma^Z_{eff}$ is more sensitive to the OOP texture, as shown by the comparison of Fig. 3(d) and Fig. 4(b) for $\theta_0 = -45°$. This suggests that OOP texture is of more importance for unconventional SOT materials featuring Z-polarized spins.

Although we only calculated the effective device SHC for Mn$_3$GaN, this approach can be used for any unconventional SOT materials based on crystal or magnetic structures. Two key observations emerge from our analysis. First, since out-of-plane (OOP) spin components ($\sigma^Z_{eff}$) are more effective for magnetization manipulation than spins polarized in other directions [20], [21], it is crucial to identify the grain orientation that maximizes $\sigma^Z$. For Mn$_3$GaN, with the axes defined in this study, this optimal in-plane angle is -45°. Therefore, one should apply the current along the $[1\bar{1}0]$ direction of Mn$_3$GaN (001) films. Second, to achieve a high USHC value, the material should be grown with a strong texture, meaning that the spread widths of the in-plane and OOP texture should be minimized as much as possible.

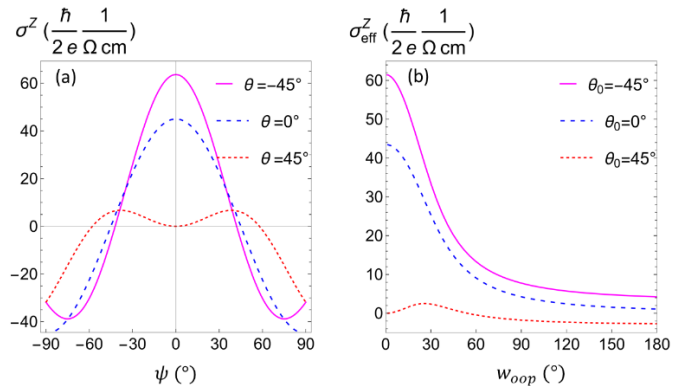

Fig. 4. (a) Computed dependence of SHC of Z-spin on the out-of-plane orientation ($\psi$) for three different $\theta_0$ values. (b) Computed effective SHC values of Z-spin as a function of the out-of-plane orientation spread width ($w_{oop}$) for Mn$_3$GaN. The in-plane orientation spread width ($w_{ip}$) is set as 15°. The calculations are based on SHC values of Mn$_3$GaN [12].

The width value of the ODF can be measured experimentally by X-ray diffraction [28]. In particular, the $\phi$-scan and $\omega$-scan quantitatively measure the in-plane and OOP texture, respectively. The full width at half maximum (FWHM) of the diffraction peak is related to width by

$$\text{FWHM} = 2\sqrt{2ln2} \text{ width} \quad (10)$$

The FWHM of a $\phi$-scan peak of sputtered materials can vary among different material systems. It can be as low as a few degrees with ion-beam assistance [29] and as broad as tens of degrees by standard normal sputtering [30]. Based on this, we choose a width values of 1° to represent highly textured or epitaxial material and 15° for textured material. For $Mn_3GaN$, the USHC shows good tolerance to in-plane spread width, as the $\sigma_{eff}^Z$ with width of 15° is still 96.7% of the value for a width of 1° when $\theta_0$ = -45° (Fig. 4(d)). In contrast, a 15° of OOP spread width will reduce the $\sigma_{eff}^Z$ to 84.6% of its maximum value.

Sputtering is an industry-friendly technique due to its high reliability and suitability for large-scale production [24], [25], [31], [32]. However, achieving single-crystal growth via sputtering can be challenging. As previously noted, the USHC can still remain high even with a slight spread in orientations, represented by a small width in the orientation distribution function (ODF). Here, we discuss three techniques that can assist sputtering in achieving highly textured material growth. The first technique is ion-beam assisted deposition (IBAD), which employs an ion beam to bombard the film during deposition, promoting crystal orientation alignment with the direction of the incident ions [29], [33]. Additionally, off-axis sputtering [34]-[36] can also improve the texture by positioning the substrate off the target axis, unlike on-axis sputtering where the target faces the substrate directly. In off-axis sputtering, only low-energy atoms reach the substrate, minimizing damage to the film. The third method leverages substrates with step edges, along which atoms tend to nucleate, resulting in a preferred orientation. This approach is also commonly used in the growth of single-crystal 2D materials [37], [38].

### III. CONCLUSION

In summary, we have analyzed the influence of polycrystalline effects on the SHC values of SOT materials. For high-symmetry materials like Pt, the conventional SHC remains unaffected by grain orientation. However, in low-symmetry materials capable of generating spin polarization along the $X$, $Y$ and $Z$ directions, both conventional and unconventional SHC depend on grain orientation. A Gaussian-shaped ODF was employed to quantitatively describe the texture of the SOT material, and the effective SHC values for devices composed of polycrystalline materials were calculated for Pt and $Mn_3GaN$. This approach can be easily extended to other SOT materials. From our analysis, we identified two essential factors for maximizing OOP spins: the choice of current injection direction and the crystallographic texture. To enhance texture or narrow the ODF width, techniques such as ion-beam assisted deposition (IBAD), off-axis sputtering, and step-edge guidance during sputtering deposition can be employed. Our study provides a guideline for maximizing the efficiency of unconventional spins, which is crucial for developing energy-efficient and ultrafast spintronics-based logic and memory devices.

### ACKNOWLEDGMENT

This work was supported, in part, by SMART, one of the seven centers of nCORE, a Semiconductor Research Corporation program, sponsored by the National Institute of Standards and Technology (NIST) and by the Global Research Collaboration (GRC) Logic and Memory program (No. 31796685), sponsored by Semiconductor Research Corporation (SRC).

### REFERENCES

[1] J. Ryu, S. Lee, K. J. Lee, and B. G. Park, 'Current-Induced Spin–Orbit Torques for Spintronic Applications', *Adv. Mater.*, vol. 32, no. 35, p. 1907148, Sep. 2020, doi: 10.1002/ADMA.201907148.

[2] J.-P. Wang *et al.*, 'A Pathway to Enable Exponential Scaling for the Beyond-CMOS Era', *Proceedings of the 54th Annual Design Automation Conference,* pp. 1–6, Jun. 2017, doi: https://doi.org/10.1145/3061639.3072942.

[3] A. Manchon *et al.*, 'Current-induced spin-orbit torques in ferromagnetic and antiferromagnetic systems', *Rev. Mod. Phys.*, vol. 91, no. 3, p. 35004, Sep. 2019, doi: 10.1103/RevModPhys.91.035004.

[4] S. Ikeda *et al.*, 'A perpendicular-anisotropy CoFeB--MgO magnetic tunnel junction', *Nat. Mater.*, p. 11, 2010, doi: 10.1038/NMAT2804.

[5] G. Yu *et al.*, 'Switching of perpendicular magnetization by spin-orbit torques in the absence of external magnetic fields', *Nat. Nanotechnol.|*, vol. 9, 2014, doi: 10.1038/NNANO.2014.94.

[6] L. You *et al.*, 'Switching of perpendicularly polarized nanomagnets with spin orbit torque without an external magnetic field by engineering a tilted anisotropy', *Proc. Natl. Acad. Sci.*, vol. 112, no. 33, pp. 10310–10315, Aug. 2015, doi: 10.1073/PNAS.1507474112.

[7] S. Fukami, C. Zhang, S. Duttagupta, A. Kurenkov, and H. Ohno, 'Magnetization switching by spin-orbit torque in an antiferromagnet-ferromagnet bilayer system', *Nat. Mater.*, p. 15, 2016, doi: 10.1038/NMAT4566.

[8] Z. Zhao *et al.*, 'External-Field-Free Spin Hall Switching of Perpendicular Magnetic Nanopillar with a Dipole-Coupled Composite Structure', *Adv. Electron. Mater.*, 2020, doi: 10.1002/aelm.201901368.

[9] S.-H. C. Baek *et al.*, 'Spin currents and spin-orbit torques in ferromagnetic trilayers', *Nat. Mater.*, vol. 17, pp. 509–513, 2018, doi: 10.1038/s41563-018-0041-5.

[10] M. Wang *et al.*, 'Field-free switching of a perpendicular magnetic tunnel junction through the interplay of spin-orbit and spin-transfer torques', *Nat. Electron.*, 2018, doi: 10.1038/s41928-018-0160-7.

[11] D. MacNeill, G. M. Stiehl, M. H. D Guimaraes, R. A. Buhrman, J. Park, and D. C. Ralph, 'Control of spin-orbit torques through crystal symmetry in $WTe_2$/ferromagnet bilayers', *Nat. Phys.*, 2017, doi: 10.1038/NPHYS3933.

[12] T. Nan *et al.*, 'Controlling spin current polarization through non-collinear antiferromagnetism', *Nat. Commun.*, 2020, doi: 10.1038/s41467-020-17999-4.

[13] A. Bose *et al.*, 'Tilted spin current generated by the collinear antiferromagnet ruthenium dioxide', *Nat. Electron.*, vol. 5, no. 5, pp. 267–274, 2022, doi: 10.1038/s41928-022-00744-8.

[14] M. DC *et al.*, 'Observation of anti-damping spin–orbit torques generated by in-plane and out-of-plane spin polarizations in $MnPd_3$', *Nat. Mater.*, pp. 1–8, 2023, doi: 10.1038/s41563-023-01522-3.

[15] Y. Liu et al., 'Field-free switching of perpendicular magnetization at room temperature using out-of-plane spins from TaIrTe$_4$', *Nat. Electro.*, vol. 6, no. 10, pp. 732–738, 2023, doi: 10.1038/s41928-023-01039-2.

[16] J.-P. Wang, T. Low, Y. Yang, and S. Lee, 'Materials generating multi spin components for magnetization switching and dynamics', *US Patent*, vol. US20240172565A1, Oct. 2023, Accessed: Jun. 29, 2024. [Online]. Available: https://patents.google.com/patent/US20240172565A1/en

[17] Y. Yang et al., 'Giant spin Hall effect with multi-directional spin components in Ni$_4$W', Nov. 2024, Accessed: Nov. 09, 2024. [Online]. Available: https://arxiv.org/abs/2411.05682v1

[18] Y. Yang et al., 'Coexistence of unconventional spin Hall effect and antisymmetric planar Hall effect in IrO$_2$', Nov. 2024, Accessed: Nov. 09, 2024. [Online]. Available: https://arxiv.org/abs/2411.05688v1

[19] A. Roy, M. H. D. Guimarães, and J. Sławińska, 'Unconventional spin Hall effects in nonmagnetic solids', *Phys. Rev. Mater.*, vol. 6, no. 4, p. 045004, 2022, doi: 10.1103/PHYSREVMATERIALS.6.045004/FIGURES/3/MEDIUM.

[20] D. J. P. De Sousa, P. M. Haney, J. P. Wang, and T. Low, 'Field-Free-Switching State Diagram of Perpendicular Magnetization Subjected to Conventional and Unconventional Spin-Orbit Torques', *Phys. Rev. Appl.*, vol. 18, no. 5, p. 054020, 2022, doi: 10.1103/PHYSREVAPPLIED.18.054020/FIGURES/7/MEDIUM.

[21] D. K. Lee and K. J. Lee, 'Spin-orbit Torque Switching of Perpendicular Magnetization in Ferromagnetic Trilayers', *Sci. Rep.*, vol. 10, no. 1, pp. 1–7, 2020, doi: 10.1038/s41598-020-58669-1.

[22] L. Zhu, D. C. Ralph, and R. A. Buhrman, 'Maximizing spin-orbit torque generated by the spin Hall effect of Pt', *Appl. Phys. Rev.*, vol. 8, no. 3, p. 031308, 2021, doi: 10.1063/5.0059171.

[23] M. Dc et al., 'Room-temperature high spin–orbit torque due to quantum confinement in sputtered Bi$_x$Se$_{(1-x)}$ films', *Nat. Mater.*, vol. 17, no. 9, pp. 800–807, 2018, doi: 10.1038/s41563-018-0136-z.

[24] D. Zhang et al., 'Robust negative longitudinal magnetoresistance and spin–orbit torque in sputtered Pt$_3$Sn and Pt$_3$Sn$_x$Fe$_{1-x}$ topological semimetal', *Nat. Commun.*, vol. 14, no. 1, pp. 1–8, 2023, doi: 10.1038/s41467-023-39408-2.

[25] Y. Fan et al., 'Observation and enhancement of room temperature bilinear magnetoelectric resistance in sputtered topological semimetal Pt$_3$Sn', *npj Spintronics*, vol. 2, no. 1, pp. 1–6, 2024, doi: 10.1038/s44306-024-00036-1.

[26] A. Van Pamel, G. Sha, and M. J. S. Lowe, 'Attenuation and Phase Velocity of Elastic Wave in Textured Polycrystals with Ellipsoidal Grains of Arbitrary Crystal Symmetry', *Acoustics*, vol. 2, no. 1, pp. 51–72, 2020, doi: 10.3390/ACOUSTICS2010005.

[27] Paritosh, D. J. Srolovitz, C. C. Battaile, X. Li, and J. E. Butler, 'Simulation of faceted film growth in two-dimensions: microstructure, morphology and texture', *Acta Mater.*, vol. 47, no. 7, pp. 2269–2281, 1999, doi: 10.1016/S1359-6454(99)00086-5.

[28] U. F. Kocks, C. N. Tomé, and H.-R. Wenk, 'Plasticity Modeling in Minerals and Rocks', *Texture and Anisotropy: Preferred Orientations in Polycrystals and Their Effect on Materials Properties*, pp. 560–596, 2000.

[29] S. Kreiskott et al., 'Reel-to-reel preparation of ion-beam assisted deposition (IBAD)-MgO based coated conductors', *Supercond. Sci. Technol.*, vol. 17, no. 5, p. S132, 2004, doi: 10.1088/0953-2048/17/5/008.

[30] S. Mahieu, P. Ghekiere, D. Depla, and R. De Gryse, 'Biaxial alignment in sputter deposited thin films', *Thin Solid Films*, vol. 515, no. 4, pp. 1229–1249, 2006, doi: 10.1016/J.TSF.2006.06.027.

[31] P. Sahu et al., 'Room Temperature Spin-to-Charge Conversion in Amorphous Topological Insulating Gd-Alloyed Bi$_x$Se$_{1-x}$/CoFeB Bilayers', *ACS Appl. Mater. Interfaces*, vol. 15, no. 32, pp. 38592–38602, 2023, doi: 10.1021/ACSAMI.3C07695.

[32] M. Kam, Q. Zhang, D. Zhang, and Z. Fan, 'Room-Temperature Sputtered SnO$_2$ as Robust Electron Transport Layer for Air-Stable and Efficient Perovskite Solar Cells on Rigid and Flexible Substrates', *Sci. Rep.*, vol. 9, no. 1, pp. 1–10, 2019, doi: 10.1038/s41598-019-42962-9.

[33] V. Matias and R. H. Hammond, 'Ion beam induced crystalline texturing during thin film deposition', *Surf. Coat Technol.*, vol. 264, pp. 1–8, 2015, doi: 10.1016/J.SURFCOAT.2014.12.018.

[34] J. C. Gallagher et al., 'Epitaxial growth of iridate pyrochlore Nd$_2$Ir$_2$O$_7$ films', *Sci. Rep. 2016 6:1*, vol. 6, no. 1, pp. 1–7, Feb. 2016, doi: 10.1038/srep22282.

[35] B. Peters et al., 'Epitaxial films of Heusler compound Co$_2$FeAl$_{0.5}$Si$_{0.5}$ with high crystalline quality grown by off-axis sputtering', *Appl. Phys. Lett.*, vol. 103, no. 16, p. 162404, 2013, doi: 10.1063/1.4825338.

[36] P. W. Swatek et al., 'Room temperature spin-orbit torque efficiency in sputtered low-temperature superconductor δ-TaN', *Phys. Rev. Mater.*, vol. 6, no. 7, p. 074206, 2022, doi: 10.1103/PHYSREVMATERIALS.6.074206/FIGURES/7/MEDIUM.

[37] Y. Wan, J. H. Fu, C. P. Chuu, V. Tung, Y. Shi, and L. J. Li, 'Wafer-scale single-orientation 2D layers by atomic edge-guided epitaxial growth', *Chem. Soc. Rev.*, vol. 51, no. 3, pp. 803–811, 2022, doi: 10.1039/D1CS00264C.

[38] Z. Zhang, X. Yang, K. Liu, and R. Wang, 'Epitaxy of 2D Materials toward Single Crystals', *Adv. Sci.*, vol. 9, no. 8, p. 2105201, 2022, doi: 10.1002/ADVS.202105201.